\documentclass{pasj00}
\begin{document}
\SetRunningHead{N. Ota, K. Onzuka, \& K. Masai}{Density Profile of a Cool Core of Galaxy Clusters} 
\Received{}
\Accepted{}

\title{Density Profile of a Cool Core of Galaxy Clusters}

\author{%
Naomi \textsc{Ota},\altaffilmark{1}
Kiyokazu \textsc{Onzuka},\altaffilmark{2}
and 
Kuniaki \textsc{Masai} \altaffilmark{2}
}

\altaffiltext{1}{Department of Physics, Nara Women's University, 
                 Kitauoyanishi-machi, Nara, Nara 630-8506}
\email{naomi@cc.nara-wu.ac.jp}
\altaffiltext{2}{Department of Physics, Tokyo Metropolitan University, 
   1-1 Minami-Osawa, Hachioji, Tokyo 192-0397}
 
\KeyWords{galaxies: clusters: general -- galaxies: cooling flows -- 
     galaxies: intergalactic medium -- X-rays: galaxies: clusters } 

\maketitle

\begin{abstract}
  The density profile of a cool core of intracluster gas is
  investigated for a cluster of galaxies that is initially in the
  virial equilibrium state, and then undergoes radiative cooling. The
  initial gas profile is derived under the assumption that the gas is
  hydrostatic within the dark-matter potential presented by the
  ``NFW'' or King model, and has a polytropic profile. The
  contribution of masses of gas and galaxies to the potential in the
  calculation is ignored compared to the dark matter. The temperature
  and density profiles of gas in its quasi-hydrostatic cooling phase,
  which is expected to last for $\sim$Gyr, is then calculated for
  different initial gas profiles.  It is found that in the
  quasi-hydrostatic cooling phase, while the temperature decreases to
  about one-third, the density increases by a factor of 4--6 at the
  cluster center in comparison with their initial polytropic values,
  though the profiles over the core depend on the dark-matter
  potential.  Hence, the core radius in the quasi-hydrostatic cooling
  gas appears to be smaller than that in the initial polytropic
  gas. We compared the density profile of the cool core with
  observations to find that, while the initial density is around the
  upper bounds of large-core ($> 100$~kpc) clusters, most likely
  relaxed, but the cooling is not yet significant, the central density
  under quasi-hydrostatic cooling falls between the mid- and
  high-values of small ($< 100$~kpc)- or cool-core clusters.  It is
  also found for the quasi-hydrostatic cooling gas that the entropy
  profile roughly agrees with the best-fit model to the ACCEPT cluster
  sample with a low central entropy; also, the pressure gradient in
  the inner core is close to that of the REXCESS sample. X-ray surface
  brightness calculated for the quasi-hydrostatic cooling gas is well
  represented by the conventional double $\beta$-model, giving a
  physical basis for applying the double $\beta$-model to cool-core
  clusters.
 \end{abstract}

\section{Introduction}\label{sec:intro}
Clusters of galaxies offer excellent laboratories for the thermal
evolution of cosmic baryons; the objects are filled with hot ($T\sim
10^7-10^8$~K), highly ionized intracluster gas, and the baryon fraction
in massive clusters is close to the cosmic value.  Since the time scale
of radiative cooling at the center of regular clusters is estimated to
be shorter than the Hubble time, their core regions are thought to be
affected by cooling.  According to statistical studies of clusters
using X-ray observations, the proportion of clusters having a compact,
cool core (often termed CC clusters) is roughly 50\% (e.g.,
\cite{ohara06,ota06,chen07,santos08,cavagnolo09,hudson10}).

Based on earlier works on the peaked surface brightness that cooling core
clusters exhibit, it was suggested that the global cooling-flow would
occur unless some heating process balances with radiative cooling
(e.g., \cite{fabian94}).  The similarity and smoothness in cooling
profiles indicate the need for continuous, distributed heat source (for
review, e.g., \cite{peterson06}).  On the other hand, a possible mild
inflow with a quasi-hydrostatic balance is proposed to account for
the observed temperature profiles of cooling cores \citep{masai04}.
Later, using a hydrodynamics code,
\citet{akahori06} demonstrate that
such a cooling phase lasts for $\sim$~Gyr, and then breaks down into
global cooling-flow.  They discuss the origin of the observed two-peaked
core-size distribution \citep{ota02,ota04} in view of the thermal
evolution.

While the intracluster gas has often been studied on the basis of the
conventional isothermal $\beta$-model, a deviation therefrom has been
commonly observed at the center of cooling-core clusters; they exhibit
a systematically higher central density, while their profiles are
fairly universal outside $0.1r_{500}\sim 100$~kpc (e.g.,
\cite{neumann99}).  To better reproduce those X-ray observations, some
authors have introduced empirical models, such as a double
$\beta$-model (e.g., \cite{jones84}), and a modified $\beta$-model
with a cool density cusp at the center and steepening at a large
radius \citep{vikhlinin06}. This model was well fitted to the
high-resolution {\it Chandra} data including the core emission (see
also \cite{bulbul10}).  Although those models are given in analytic
forms, and therefore are simply utilized for observational studies,
the physical background that they are based on is not always clear.

In the present paper, we intend to obtain a fiducial-density profile
of cooling gas under quasi-hydrostatic balance.  \citet{masai04} give
a temperature profile for the {\it initially} isothermal gas, but no
density profile.  We calculate the temperature of cooling gas in the
same manner, but more generally consider that the gas is initially
hydrostatic with an polytropic profile in the NFW \citep{navarro97} or
King dark matter potential.  In section~2, the calculation of our
model is described, and pressure and entropy as well as density of
quasi-hydrostatic cooling gas are presented.  In section~3, we compare
those calculated quantities with observations, and also examine the
application of the widely used $\beta$-model to X-ray surface
brightness calculated for quasi-hydrostatic cooling gas.  We then
discuss implications on the intracluster medium (ICM) thermal
evolution in the cluster core regions.  Since the profiles of gas
density etc. of cool cores are of interest to us, we present the
quantities normalized by their central values unless specified
otherwise.

\section{Calculations}

\subsection{Initial Hydrostatic Gas Distribution}\label{subsec:initial}

\begin{figure}
\begin{center}
\FigureFile(60mm,80mm){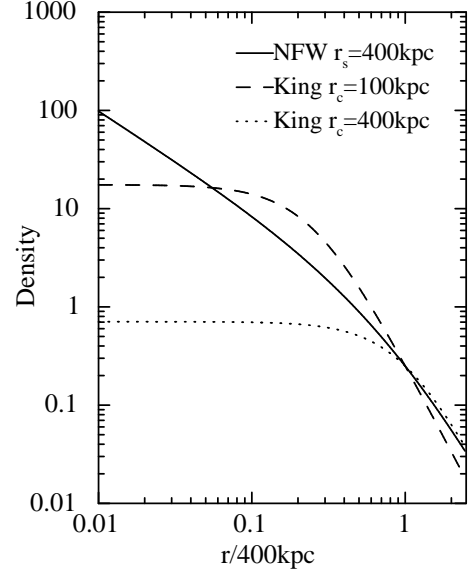}
\end{center}
\caption{Dark-matter density profiles with $r_{\rm s} = 400$~kpc for
  the NFW (solid) and $r_{\rm c} = 100$~kpc for the King (broken)
  distribution taken into consideration here (see text).  The King
  profile with $r_{\rm c}=400$~kpc is also shown for reference.  The
  three curves are normalized so as to pass the same point at
  $r=400$~kpc.  }\label{fig:potential}
\end{figure}

We consider a spherically symmetric galaxy cluster initially in the
virial equilibrium.  We ignore the mass of the gas and galaxies
compared to the dark matter, which is thought to occupy typically more
than 80\% of the cluster mass.  Thus, the gas distributes so that its
pressure balances with the local gravitational potential formed by the
dark matter.  For a noncool core (NCC), i.e, the initial state
before the cooling becomes significant, we assume that the gas is in a
hydrostatic balance,
\begin{equation}
{1 \over \rho}{d P \over d r} = - {d \phi \over d r} \simeq - G {M_{\rm DM}(r) \over r^2}\ ,
\end{equation}
where $P$ and $\rho$ are the gas pressure and the gas density, respectively,
and $\phi$ the gravitational potential; $M_{\rm DM}(r)$ is the dark-matter mass 
contained in a radius $r$.  We solve equation (1) with an
equation of state as $P \propto \rho^\gamma$ to obtain the initial-gas
distribution, where $\gamma = 1+(1/N)$ for the polytrope index $N$.  We
consider the NFW distribution and the approximated King distribution
(referred to as King distribution hereafter) for the dark-matter
potential as shown in figure~\ref{fig:potential}.

The gas-temperature profile thus obtained for the NFW potential can be
expressed in the form
\begin{equation}
{T(r^*) \over T(0)} = 1 - {\gamma-1 \over \gamma} {U_{\rm G} \over kT(0)} \left[ 1-{\log(1+r^*) \over r^*} \right]
\end{equation}
with
\begin{equation}
r^* = {r \over r_{\rm s}} \quad \mbox{and} \quad U_{\rm G} = {\mu m G \over r_{\rm s}} 4\pi r_{\rm s}^3 \delta_{\rm c} \rho_{\rm c}\ ,
\end{equation}
where $r_{\rm s}$ is the characteristic radius of the NFW distribution, expressed as
\begin{equation}
\rho_{\rm DM} (r^*) = {\delta_{\rm c} \rho_{\rm c} \over r^* (1+r^*)^2}\ ,
\end{equation}
and $\delta_{\rm c} \rho_{\rm c}$ is a constant related to the dark-matter mass, $M_{\rm DM}$, through 
\begin{equation}
M_{\rm DM}(r^*) = 4\pi r_{\rm s}^3 \delta_{\rm c} \rho_{\rm c} \left[ \log (1+r^*) + {1 \over 1+r^*} -1 \right]\ .
\end{equation}
The gas-density profile is given by
\begin{equation}
{\rho(r^*) \over \rho(0)} = \left[ {T(r^*) \over T(0)} \right]^{1/(\gamma -1)} = \left[ {T(r^*) \over T(0)} \right]^N\ .
\end{equation}

Similarly we express the gas temperature for the King potential as
\begin{equation}
{T(r^*) \over T(0)} = 1 - {\gamma -1 \over \gamma} {U_{\rm G} \over kT(0)} \left[ 1-\frac{\log(r^*+\sqrt{r^{\ast 2}+1})}{r^*} \right]
\end{equation}
with
\begin{equation}
r^* = {r \over r_{\rm c}} \quad \mbox{and} \quad U_{\rm G} = {\mu m G \over r_{\rm c}} 4\pi r_{\rm c}^3\rho_0\ .
\end{equation}
Here, $r_{\rm c}$ is the core radius of the King distribution, expressed as
\begin{equation}
\rho_{\rm DM} (r^*) = {\rho_0 \over (1+r^{\ast 2})^{3/2}}\ ,
\end{equation}
and $\rho_0$ is the dark-matter density at the center.  The dark-matter mass is given by
\begin{equation}
M_{\rm DM}(r^*) = 4\pi r_{\rm c}^3 \rho_0 \left[ \log (r^*+\sqrt{r^{\ast 2}+1}) - {r^* \over \sqrt{r^{\ast 2}+1}} \right]\ .
\end{equation} 
The gas-density profile for the King potential is given by equation
(6) as well.

\begin{figure*}
\begin{center}
\FigureFile(70mm,80mm){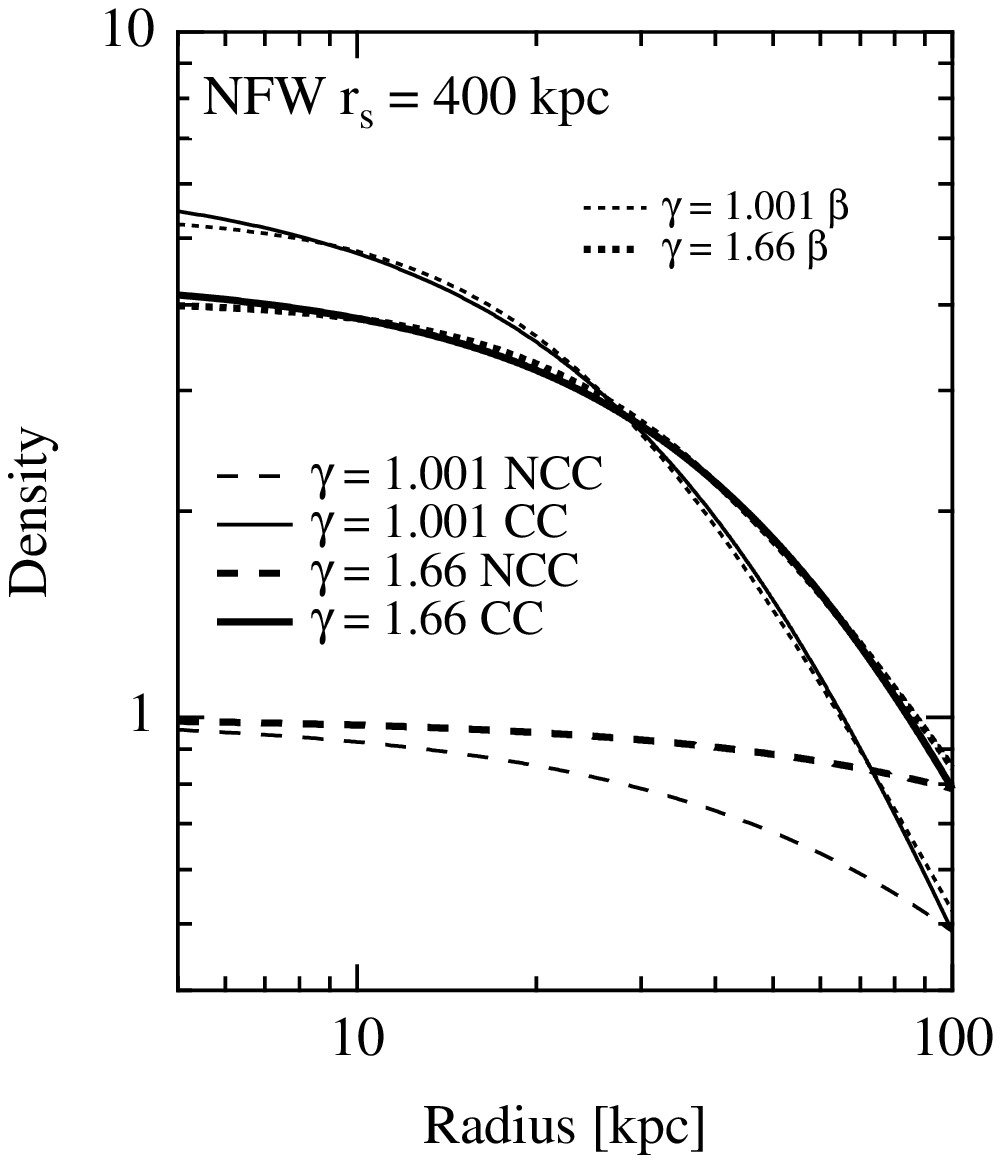}
\FigureFile(70mm,80mm){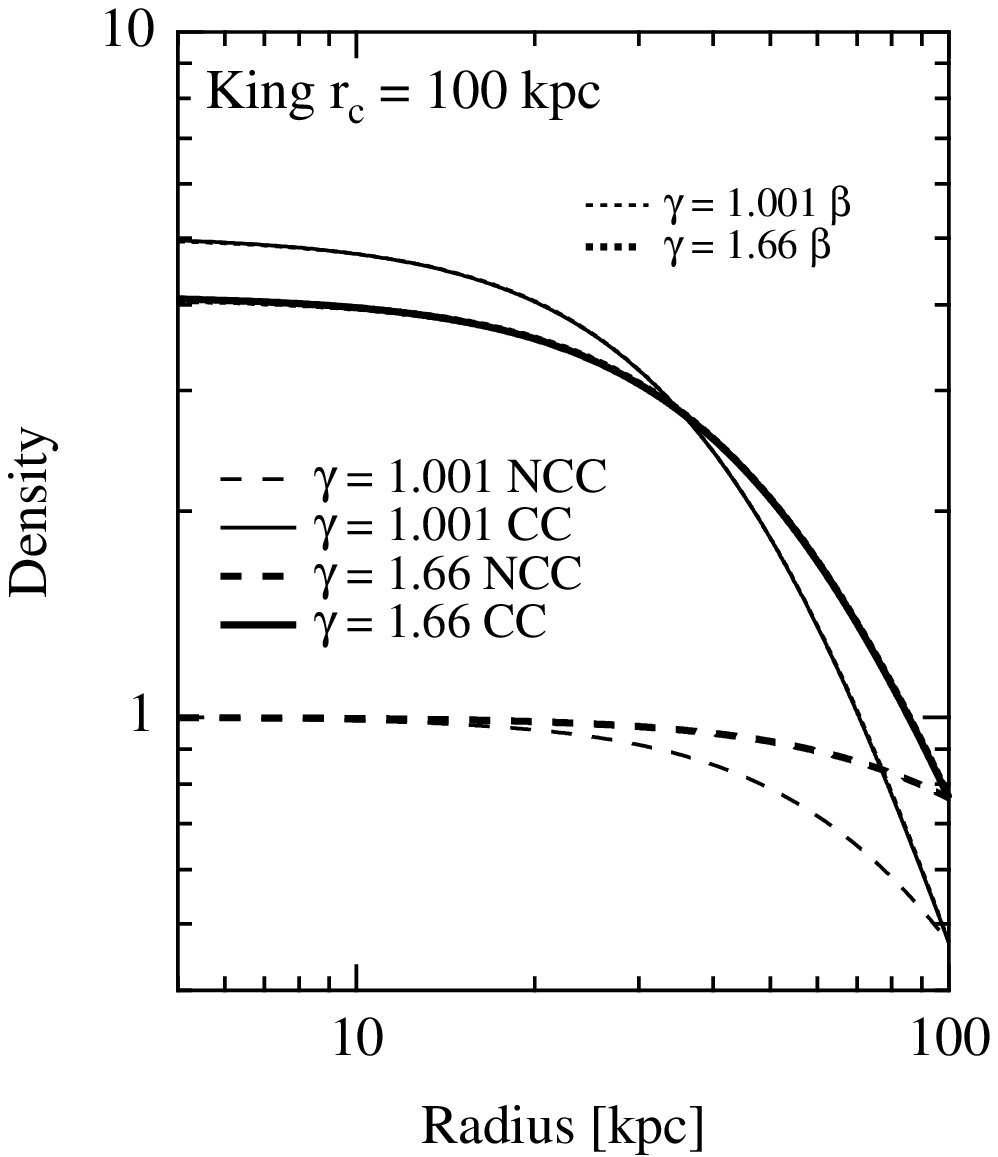}
\end{center}
\caption{Gas-density profile of quasi-hydrostatic cool cores,
  $\rho'(r)/\rho(0)$ (solid lines, CC) with $r_{\rm cool} = 100$~kpc
  and the initial noncool ones, $\rho(r)/\rho(0)$ (broken lines, NCC)
  of $\gamma =$ 1.66 (thick lines) or 1.001 (thin lines)
  polytrope. The dotted lines represent the best-fit $\beta$-models 
  (see text).  Particularly in the right panel, the best-fit  $\beta$-models (thin- and thick-dotted lines) reproduce the calculated CC density profiles (thin- and thick-solid lines) well and almost perfectly overlap with them for both $\gamma=1.001$ and $1.66$, respectively.  }\label{fig:density}
\end{figure*}

The density profiles, $\rho(r)/\rho(0)$, calculated from the above
equations are shown by the broken lines (denoted by NCC) in
figure~\ref{fig:density} with $r_{\rm s} = 400$~kpc for the NFW potential and with 
$r_{\rm c} = 100$~kpc for the King dark-matter one.  These
radii are chosen for the $\beta$-model core radius of surface
brightness ($\propto \rho^2 T^{1/2}$) to be 
100--200~kpc.  Equations (2)--(6) are equivalent to
equations (12) and (42)--(46) in \citet{suto98}, and therefore the broken
lines in the left panel (NFW NCC) of figure~\ref{fig:density}
are the same as their results.  We calculate the temperature profile,
$T(r)/T(0)$, for calculations in the following subsection as well.

\subsection{Quasi-Hydrostatic Gas Distribution under Radiative
  Cooling}\label{subsec:calc_qhs}

Now we consider the gas distribution of a cool core (CC) that
significantly undergoes radiative cooling.  The gas is then no longer
in hydrostatic balance, but should be described by hydrodynamics,
since inflow toward the inner region where the cooling rate is higher
would occur so as to compensate the pressure decrease due to radiation
loss \citep{fabian94}.  Hence, the gas-temperature and gas-density
profiles of the cluster core deviate from those polytropic profiles
given in the last subsection.

In the early cooling phase, however, inflow is still mild and a 
quasi-hydrostatic condition is attained marginally until the initial
cooling time, $\tau_{\rm cool}$. Here, $\tau_{\rm cool}$ is defined by 
the density and temperature at the center of a virialized cluster
before cooling and is typically $\sim$~Gyr \citep{akahori06}.  We
consider that the pressure decrease with radius due to radiation is
compensated immediately by the gas inflow from the adjacent outer (and
hotter) region, or the local-inflow rate is controlled by the local
cooling rate, and thus a quasi-hydrostatic condition is attained.

The profile of the cooling-gas temperature, $T'(r)$, is then expressed approximately in a form \citep{masai04}
\begin{equation}
{d \ln T' \over d \ln r} \simeq {9 \over 5} \left( 1 - {1 \over 3}\beta_T {T_0 \over T'}\right)
\end{equation}
at $0 < r \leq r_{\rm cool}$, when the variation of $T(r)$ (NCC) is
sufficiently small in this range; it should be noted that this
equation was obtained originally for an isothermal mass distribution
of the virial temperature $T_0$.  Here, $r_{\rm cool}$ is a
cooling-core radius in which the local-cooling time is shorter than
the Hubble time, as $t_{\rm cool}(r) = 3kT(r)/[2\rho(r)\Lambda] <
H_0^{-1}$ at $r < r_{\rm cool}$ and $t_{\rm cool}(r_{\rm cool}) =
H_0^{-1}$, where $\rho\Lambda$ represents the radiative-cooling rate
per unit time.  We assume $\beta_T = 1$, i.e., the gas and the dark
matter have the same temperatures.  In observations, the values of
$\beta$ implied by the density profile fall around 0.6 less than
unity.  This may, however, be a result inherent in the $\beta$-model,
because it is normally applied to observed profiles within a radius
significantly smaller than the viral radius (see \cite{akahori05}).
We put $T(0) = T_0$ for normalization and calculate $T'(r)/T(0)$ with
a boundary condition $T'(r_{\rm cool}) = T(r_{\rm cool})$.  It should
be noted that $T(r_{\rm cool})/T(0)$ depends on the potential profile
and $\gamma$, and thus $T'(r)/T(0)$ does as well, which is not
explicitly seen in equation (11), though.

Once the gas inflow takes place and attains to a steady state, and the
quasi-hydrostatic condition is achieved, we may have
\begin{equation}
{1 \over \rho'}{d P' \over d r} \simeq {1 \over \rho}{d P \over d r} \simeq - G {M_{\rm DM}(r) \over r^2}\ ,
\end{equation}
where $\rho'$ and $P'$ are the gas density and the gas pressure, respectively, of the cool core.  
The ram pressure is negligible compared to the thermal one in this balance. 
Thus, for a given $T'(r)$, the density profile
\begin{equation}
{d \ln \rho' \over d \ln r} \simeq - {d \ln T' \over d \ln r} + {T \over T'} \gamma {d \ln \rho \over d \ln r}
\end{equation}
follows; note that $\gamma$ is defined for $P$ (or $T$) and $\rho$. 
In the cool core concerned in these calculations, $T/T'$ varies with the radius from 1
at $r=r_{\rm cool}$ to $\sim 3$ at $r \sim 0$, and the gas is kept
(quasi-) hydrostatic through the core.
Actually, in order that the inflow occurs, the left-hand
side must be once larger than the right-hand side in equation (12); note that $dP/dr$ is negative.

\begin{figure*}
\begin{center}
\FigureFile(76mm,80mm){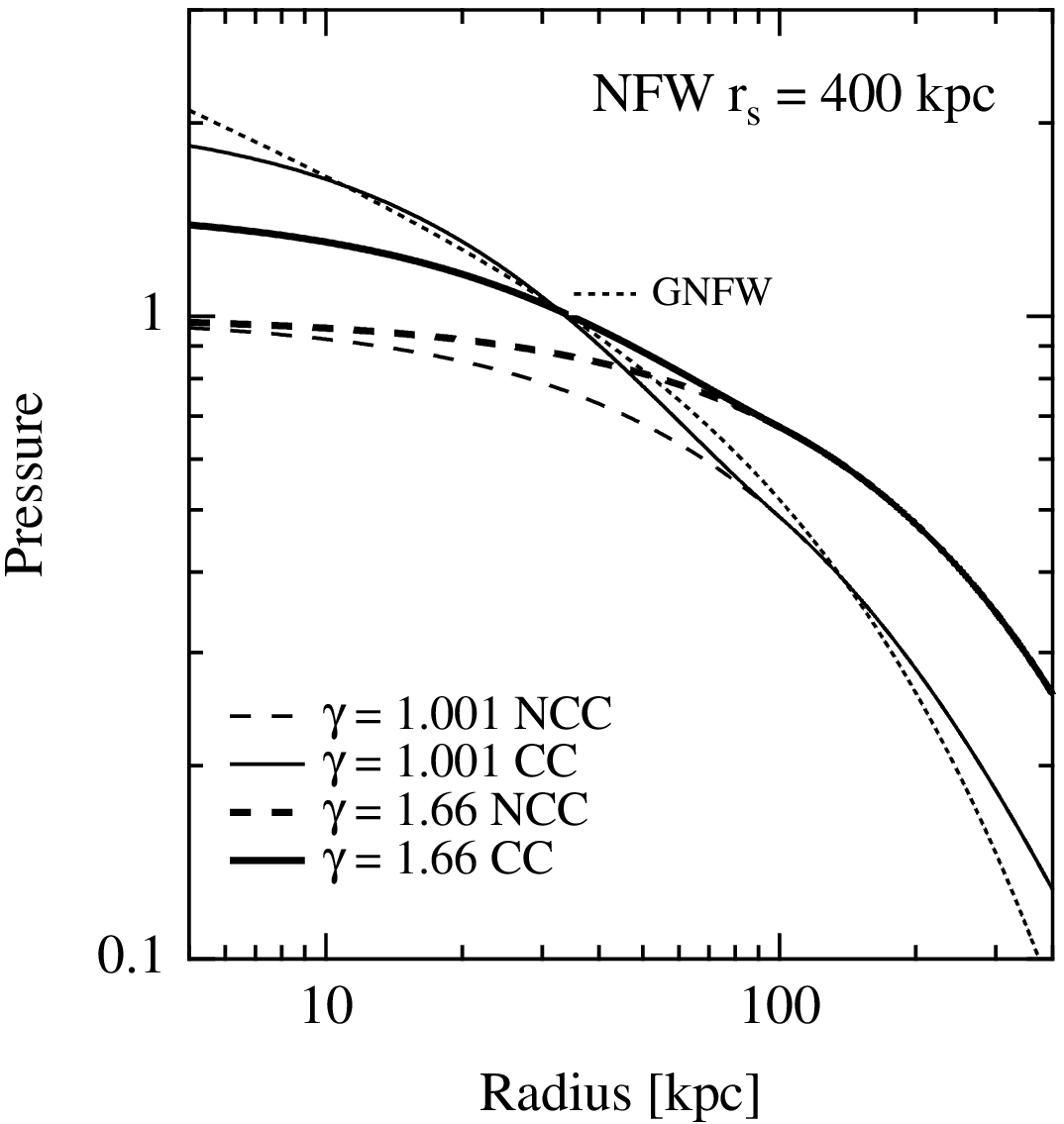}
\FigureFile(76mm,80mm){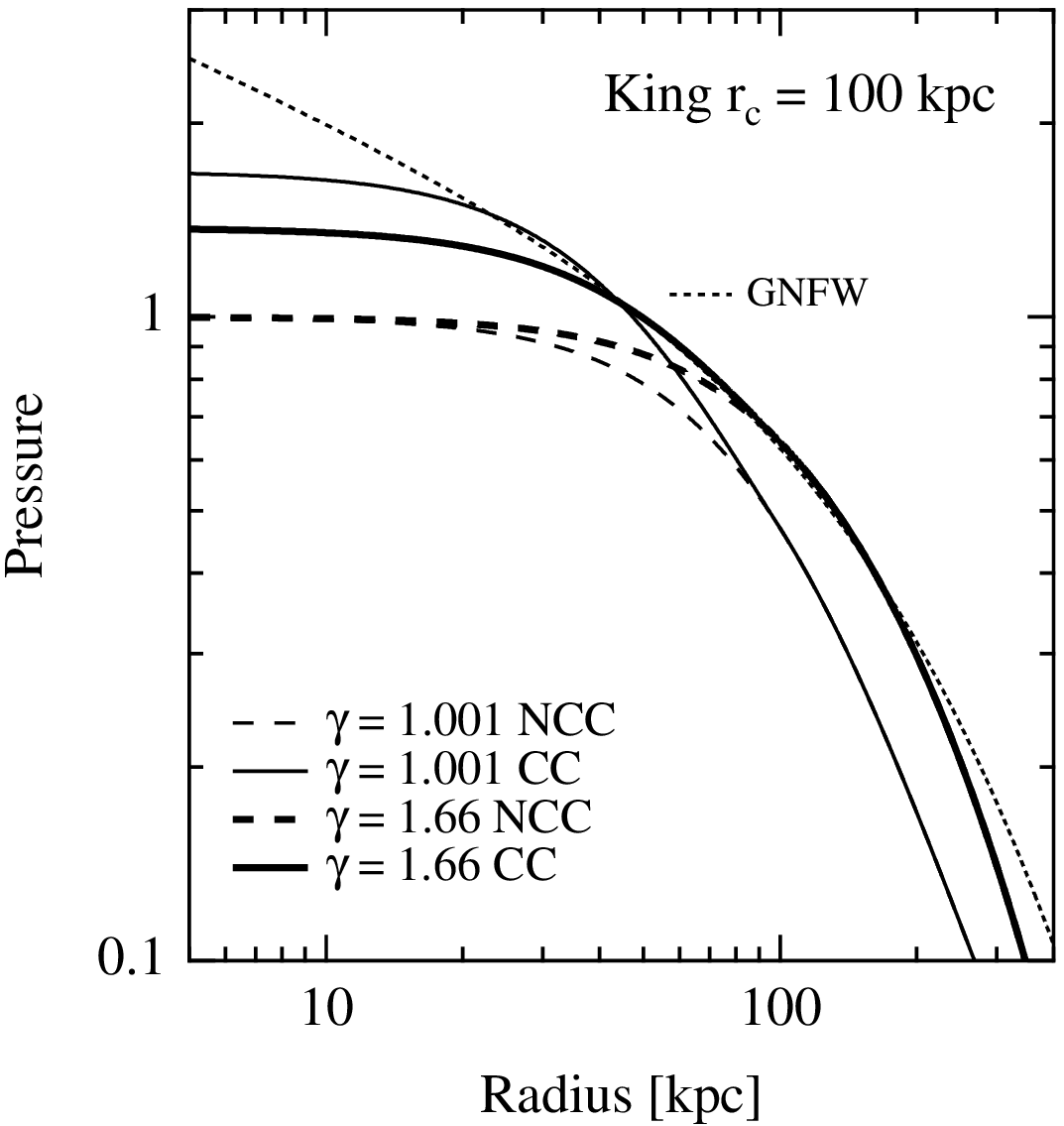}
\end{center}
\caption{Gas-pressure profile of quasi-hydrostatic
  cool cores, $\rho'(r)T'(r)/[\rho(0)T(0)]$ (solid lines, CC) with $r_{\rm cool} = 100$~kpc and the
  initial noncool ones, $\rho(r)T(r)/[\rho(0)T(0)]$(broken lines, NCC) of $\gamma =$ 1.66
  (thick lines) or 1.001 (thin lines) polytrope.  The dotted lines
  represent the GNFW-pressure profile obtained for REXCESS sample by
  \citet{arnaud10} (see text). }
\label{fig:pressure}
\end{figure*}

\begin{figure*}
\begin{center}
\FigureFile(70mm,80mm){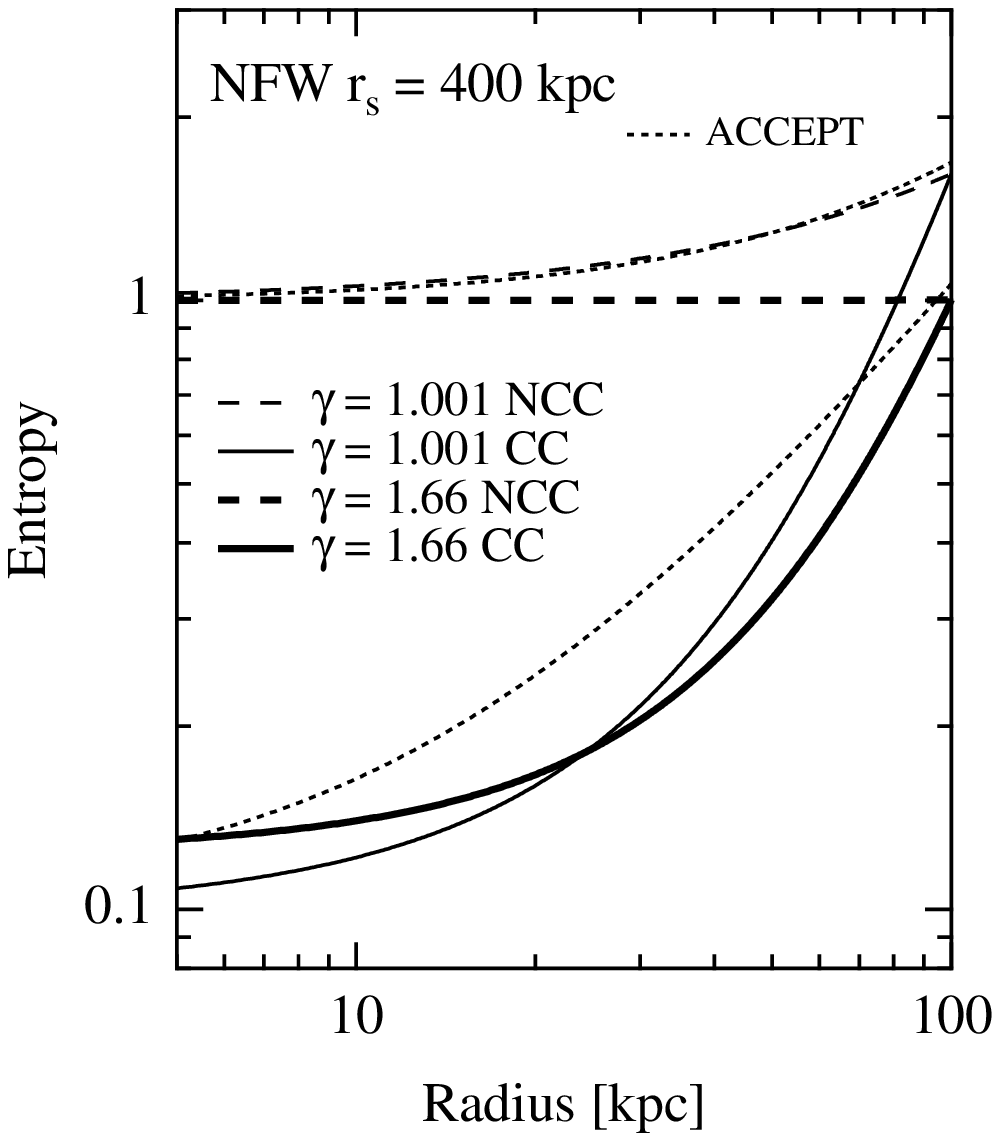}
\FigureFile(70mm,80mm){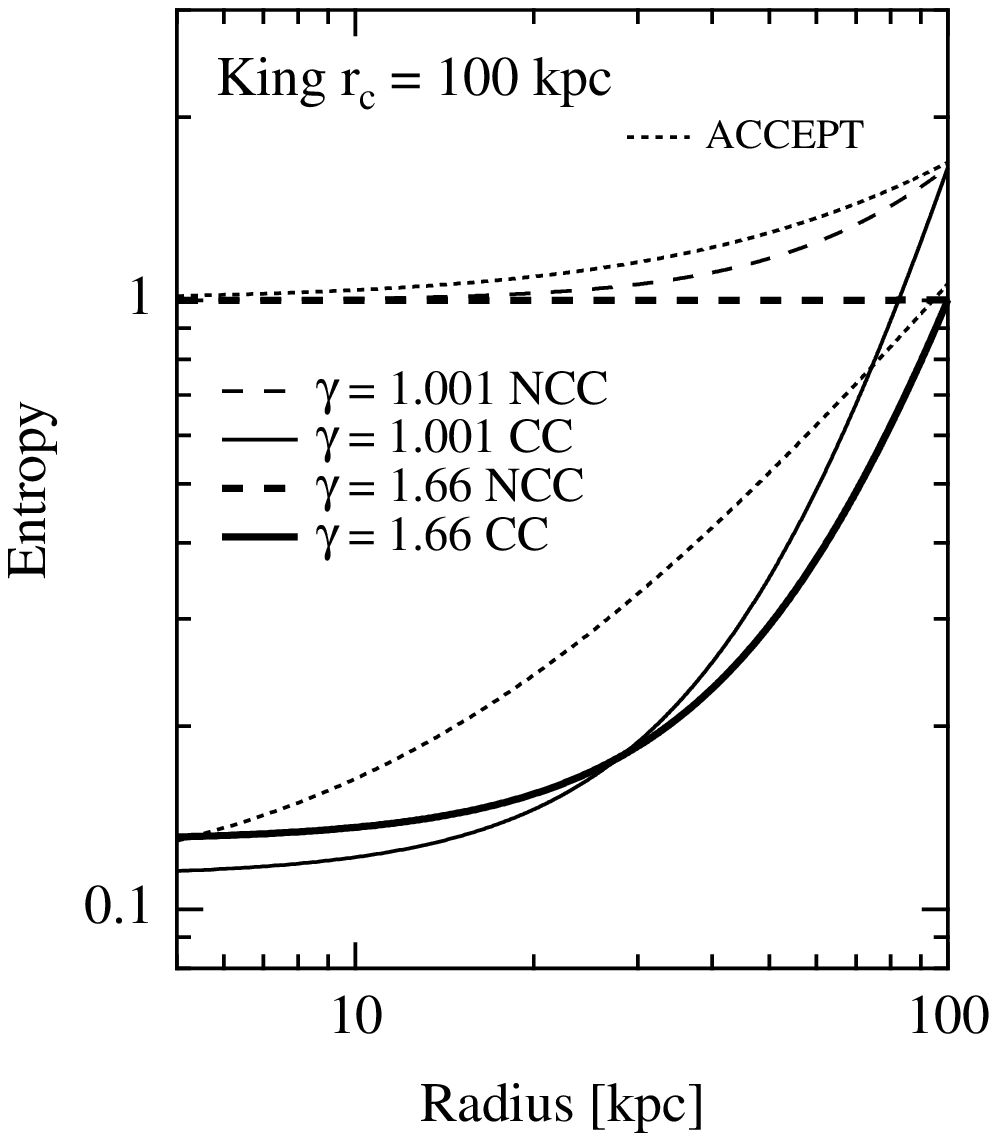}
\end{center}
\caption{Entropy profile of quasi-hydrostatic cool cores,
  $T'(r)/\rho'^{2/3}(r)/[T(0)/\rho^{2/3}(0)]$ (solid lines, CC) with
  $r_{\rm cool} = 100$~kpc and the initial noncool ones,
  $T(r)/\rho^{2/3}(r)/[T(0)/\rho^{2/3}(0)]$(broken lines, NCC) of
  $\gamma =$ 1.66 (thick lines) or 1.001 (thin lines) polytrope. The
  upper- and lower-dotted curves in each panel indicate the best-fit
  entropy profiles to the ACCEPT sample with high ($K_0>50~{\rm keV\,cm^2}$) and low ($K_0\leq50~{\rm
    keV\,cm^2}$) central entropy values \citep{cavagnolo09}, respectively.}
\label{fig:entropy}
\end{figure*}

\begin{table*}[htbp]
\caption{$\beta$-model parameters for density profiles of non-cool and cool cores.}
\begin{center}
\begin{tabular}{llllllll}\hline\hline
             & & \multicolumn{3}{c}{$\gamma=1.001$} & \multicolumn{3}{c}{$\gamma=1.66$}  \\\cline{3-5}\cline{6-8}
Potential & & $n_0$\footnotemark[*]  & $r_{\rm c,gas}$~[kpc] & $\beta$ & $n_0$ & $r_{\rm c,gas}$~[kpc] & $\beta$ \\ \hline 
NFW   	& NCC  & 0.89 (1.00) &       64 &     0.33 & 0.95 (1.00) &      104 &     0.25 \\
NFW    	& CC  & 5.41 (6.24) &       25 &     0.55 & 4.05 (4.43) &       33 &     0.45  \\ \hline
King   	& NCC & 1.00 (1.00) &      100 &     0.53 & 0.99 (1.00) &      164 &     0.44 \\
King   	& CC  & 5.02 (5.05) &       51 &     1.0 & 4.10 (4.13) &       58 &     0.81 \\ 
\hline
\end{tabular}
\end{center}
\footnotemark[*] {Central density normalized by that for the non-cool core. The value in parenthesis denotes the central surface brightness obtained from the present calculation. }

\label{tab:betafit}
\end{table*}%

Using equation (13) with $T(r)$ and $\rho(r)$ obtained in the last
subsection and $T'(r)$ from equation (11), we calculate
$\rho'(r)/\rho(0)$ with $\rho'(r_{\rm cool}) = \rho(r_{\rm cool})$,
and show it in figure~\ref{fig:density} by the solid lines (denoted by
CC) for a typical cooling radius of $r_{\rm cool} = 100$~kpc (e.g.,
\cite{peterson06}).  The initial-gas density is taken to be a
polytropic distribution of $\gamma =$ 1.001 or 1.66 (see subsection
2.1) and represented by the broken line (NCC).  The virialized gas of
$\gamma \simeq 1$ polytrope may be attained, e.g., if in the history
of the cluster formation merger shocks are radiative or thermal
conduction works in the core.  Note that the quasi-hydrostatic cooling
\citep{masai04} predicts a steeper profile than the hydrostatic one of
$\gamma = 1.001$ polytrope.

In figure~\ref{fig:density}, the dotted lines (denoted by $\beta$)
represent the best-fit $\beta$-model, $n(r) = n_0\,[ 1+ (r/r_{\rm
  c,gas})^2]^{-3\beta/2}$ with $n$ and $r_{\rm c,gas}$ being the
density and the core radius of gas, to imitate the density of CC at $5
< r/{\rm kpc} < 100$.  These parameters are given in
table~\ref{tab:betafit}, where is also shown the parameters for the
initial noncool cores in the range $5<r/{\rm kpc}<400$.  At $r <
r_{\rm cool}$ the $\beta$-model can reproduce the density of CC fairly
well for the NFW potential and quite well for the King one;
systematically, the $\beta$-model gives a flatter profile toward the
center than the calculation for the NFW.  This result may suggest that
analysis by the $\beta$-model is also useful for cool cores if the gas
is quasi-hydrostatic, though not relevant to its original meaning.

Figure~\ref{fig:pressure} shows the pressure profile, $\rho(r)
T(r)/[\rho(0) T(0)]$ (broken lines, NCC) and $\rho'(r) T'(r)/[\rho(0)
T(0)]$ (solid lines, CC).  It can be seen that in the
quasi-hydrostatic cooling core ($r \lesssim r_{\rm cool} = 100$~kpc)
the pressure profile for the initially isothermal ($\gamma = 1.001$)
gas is steeper than that for the $\gamma = 1.66$ gas, and the profile
in the inner core for the NFW potential than that for the King one.
The best-fit generalized NFW (GNFW, hereafter; \cite{nagai07})
pressure obtained for the REXCESS sample of clusters (\cite{arnaud10}
Eqs.~11--12), $p(x)=P_0 (c_{500}x)^{-\theta} [ 1 + (c_{500}x)^{\zeta}
]^{-(\eta-\theta)/\zeta}$ for $(P_0,
c_{500},\theta,\zeta,\eta)=(8.403h_{70}^{-3/2},
1.177,0.3081,1.0510,5.4905)$. The GNFW profile is normalized so as to
pass the intersection of $\gamma = 1.001$ and $\gamma = 1.66$ curves
of quasi-hydrostatic cooling gas (CC).

Figures~\ref{fig:entropy} shows the entropy profile,
$T(r)/\rho^{2/3}(r)/[T(0)/\rho^{2/3}(0)]$ (broken lines, NCC) and
$T'(r)/\rho'^{2/3}(r)/[T(0)/\rho^{2/3}(0)]$ (solid lines, CC). While
the initial polytropic gas (NCC) exhibits a flat or shallower profile,
the cooling gas (CC) exhibits entropy roughly $\propto r$ at $0.1 <
r/r_{\rm cool} < 1$ for both the NFW and King potentials.  The
best-fit profiles obtained by \citet{cavagnolo09} for the ACCEPT
cluster sample are also shown by the dotted lines, so that their
high central entropy $K_0$ ($> 50$ keV cm$^2$) is unity at $r =
0$. Note that their entropy profile consists of a power-law and a
constant, $K(r) = K_0 + K_{100}(r/100~{\rm kpc})^{\alpha}$, and the
best-fit parameters for clusters with $K_0 \leq 50$ and $K_0 > 50~{\rm
  keV\,cm^2}$ are $(K_0,K_{100},\alpha)=(16.1, 150, 1.20) and (156,
107, 1.23)$, respectively.

Comparisons between the calculated density, pressure, and entropy and
their observations are discussed in the next section. It should be
noted here that $r_{\rm cool}$ is a point at which the temperature,
and thus quantities derived therefrom, are not smoothly connected to
those on the outside, since we consider the cooling within $r_{\rm
  cool}$ while leaving the initial polytropic value as it is outside.

The picture discussed in this section is valid as long as the sound
crossing time is sufficiently shorter than $\tau_{\rm cool}$.  In
other words, without any heating process, the sound speed decreases
with the cooling, and eventually ceases to readjust the pressure
against the gravitational force.  A flattening of $\rho'(r)$ toward
the center is due to a (quasi-) hydrostatic balance, and the actual
$\rho'(r)$ could be steeper in hydrodynamics. Note that $T'$ or
$\rho'$ here is not applicable to the deep core in which the gas
accumulates, and also recall that we ignore the gas mass in comparison
to the gravitational potential.  In the quasi-hydrostatic cooling
phase, equation (12) or (13) can be a good approximation of typical
cool cores.  The results presented here are useful for the
$\beta$-model based or such a hydrostatic-balance based analysis, as
discussed in the following section.

\section{Discussion}\label{sec:discuss}
\subsection{Comparison with Observed Core Radii and Density
  profiles}\label{subsec:sb}

As can be seen in figure~\ref{fig:density} or table~\ref{tab:betafit},
the core radius in the quasi-hydrostatic cooling phase is about
one-third of the initial radius of polytropic gas in the virial
equilibrium.  If clusters that are relaxed or virialized but not yet
cooling-affected, have $\sim 200$~kpc cores, the clusters at the
quasi-hydrostatic cooling stage appear to have $\sim 60$~kpc cores.
This result is consistent with the hydrodynamical calculation by
\citet{akahori06}, who analyzed the calculated clusters using the
$\beta$-model.

\begin{figure*}[htbp]
\begin{center}
\FigureFile(80mm,80mm){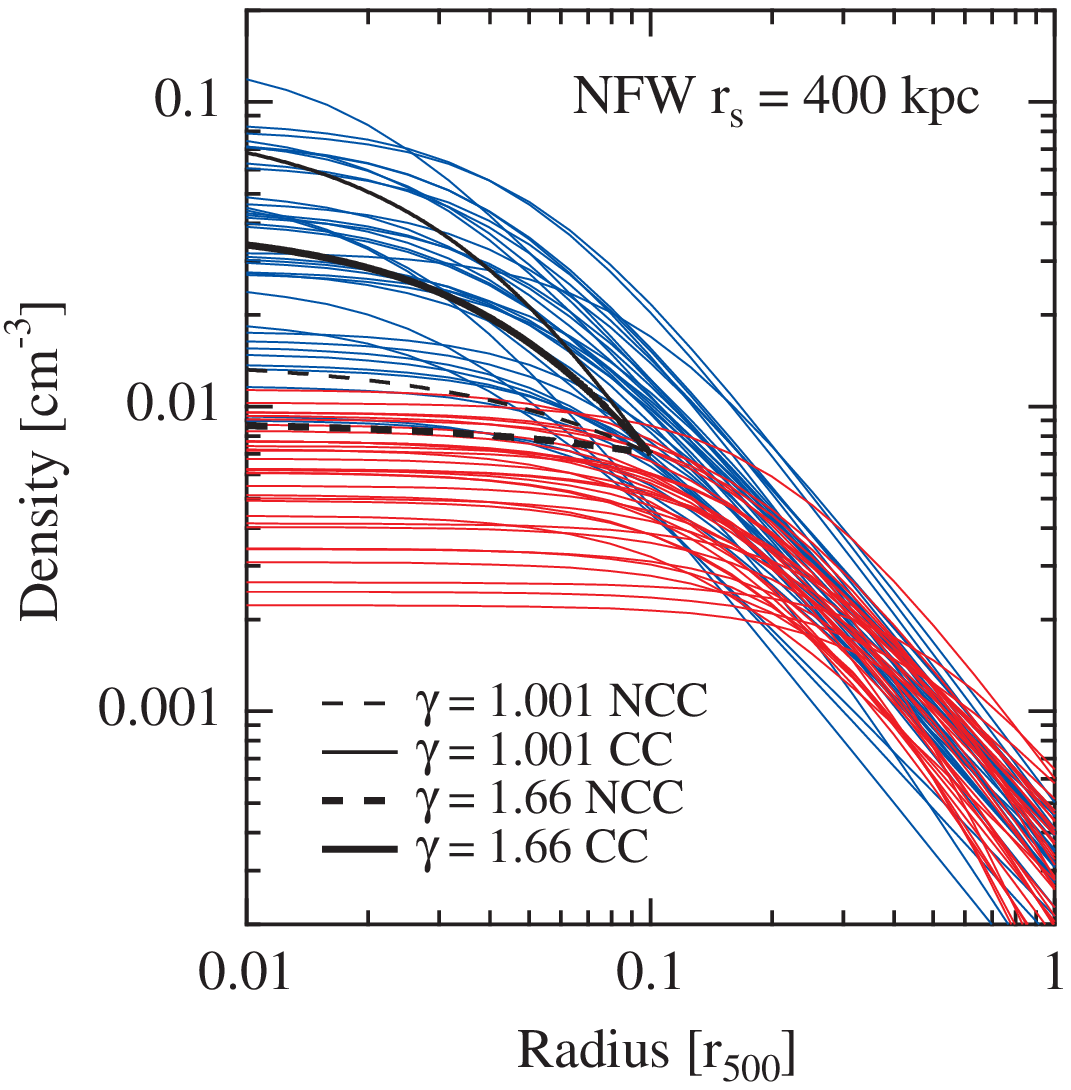}
\FigureFile(80mm,80mm){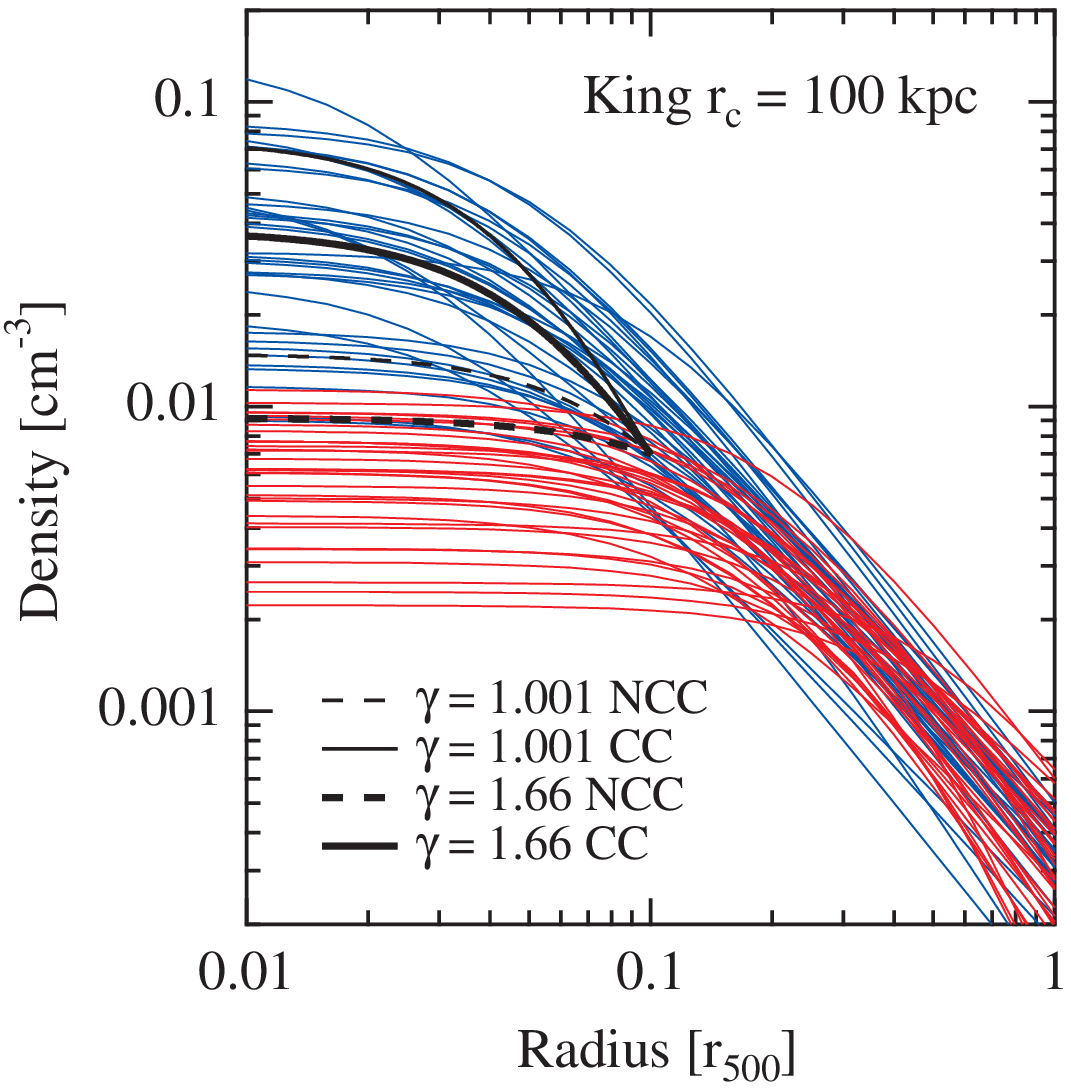}
\end{center}
\caption{Comparison of calculated gas density profiles with
  observations.  The solid and broken black lines represent the
  density profiles of quasi-hydrostatic cool cores (CC) and the
  initial non-cool cores (NCC), respectively, for the NFW (left) and
  King (right) potential.  The calculated density profiles are
  normalized such that the cooling time at $r_{\rm cool}$ is equal to
  10~Gyr.  The blue and red lines show the electron density profiles
  of clusters with small ($r_{\rm c}<100$~kpc) and large ($>$100~kpc)
  core radii, respectively (taken from the $\beta$-model analysis in
  \cite{ota06}.)}\label{fig:ne}
\end{figure*}

In figure~\ref{fig:ne}, the gas-density profiles are compared with the
best-fit $\beta$-model profiles for clusters analyzed in
\citet{ota06}.  Note that the systematic error of the central-gas
density derived under the isothermal $\beta$-model is estimated to be
small (the Appendix), which does not affect the present discussion.
In figure~\ref{fig:ne}, the solid (CC) and broken (NCC) black lines
represent the density $\rho'$ of the quasi-hydrostatic cool core and
$\rho$ of the initial polytropic noncool core, respectively.  These
calculated-density profiles are normalized such that the cooling time
at $r_{\rm cool}$ is equal to 10~Gyr in the figure.\footnote{The
  temperature at the cooling radius is assumed to be $kT(r_{\rm
    cool})= kT'(r_{\rm cool})=5$~keV in the calculation of cooling
  time, whose dependence on $kT$ is, however, so small that it does
  not affect the present discussion.}

One can see that the initial density, $\rho$ (broken lines), falls
around the upper bounds of the densities of larger-core ($r_{\rm c} >
100$~kpc) clusters, most likely relaxed, but not yet cool-core
clusters.  The density $\rho'$ (solid lines) of the quasi-hydrostatic
cool core, on the other hand, lies between the mid- and high-values of
the densities of small-core ($r_{\rm c}<100$~kpc) clusters.  We also
found from the literature on the spatially resolved X-ray spectroscopy
of clusters (\cite{allen01}; \cite{bonamente06}; \cite{zhang07};
\cite{zhang08}; \cite{snowden08}) that 11 small-core clusters out of
13 plotted in figure~\ref{fig:ne} have a clear decrease in temperature
at the center, while none of 17 large-core clusters has such a
temperature drop.

The above facts may support that the density increase in the cool core
can be accounted for by the quasi-hydrostatic cooling, and that our
calculation gives a typical density of cool cores.  However, there
exists a large scatter, and some observed clusters have higher central
densities than predicted ones.  In the context concerned here, the
scatter is ascribed to that in the potential profile and/or to those
in the initial-inflow condition and the elapsed time.

\subsection{Comparison with Observed Pressure and Entropy
  Profiles}\label{subsec:pressure}

\citet{arnaud10} discussed a universal-pressure profile for the
REXCESS cluster sample, and obtained the best-fit profile [their
equations 11--12)] based on GNFW by \citet{nagai07}.  It can be seen
from figure~\ref{fig:pressure} that the best-fit GNFW profile is close
to the overall ($10 \lesssim r/{\rm kpc} \lesssim 200$) slope of
quasi-hydrostatic cooling gas of initially $\gamma = 1.001$ polytrope
under the NFW potential.  When compared with the GNFW, a depression in
the midpart could arise from non smooth connection in $\rho'$ and $T'$
across $r_{\rm cool}$ (see subsection 2.2).  On the other hand, the
difference at $\gtrsim 200$~kpc is likely due to a difference in the
dark-matter potential or $r_{\rm s}$, or to that the cluster gas is
not simply hydrostatic.  In all cases of $\gamma$, it is unlikely that
the hydrostatic gas like the initial polytrope here (NCC), can account
for a steep profile in the inner core such as the GNFW.

\citet{cavagnolo09} derived radial entropy profiles of ICM for 239
clusters from the {\it Chandra} data (the ACCEPT sample) to find that
the model consisting of a power-law plus a constant $K(r) = K_0 +
K_{100}(r/100~{\rm kpc})^{\alpha}$ is well fitted to most entropy
profiles (see also subsection 2.2).  They also showed that the
distribution of central entropy $K_0$ is bimodal, which peaks at
$K_0\sim15$ and $\sim150~{\rm keV\,cm^2}$.  Comparing the calculated
entropy lines with the best-fit models to the ACCEPT sample
(figure~\ref{fig:entropy}), we find that the model for clusters with
the higher central entropy ($K_0>50~{\rm keV\,cm^2}$) is close to the
initial entropy profile for $\gamma=1.001$.  As for $K_0 \leq 50~{\rm
  keV\,cm^2}$, the model is reasonably in agreement with the CC cases
in the central decrement, which is lower than that of the initial NCC
profiles by about an order of magnitude.

Moreover, for the relation between the scaled temperature and the
density, \citet{arnaud10} found that the two deviations from their
average scaled profiles in the core, $r/R_{500} < 0.2$, negatively
correlate with each other (figure~3 in their paper); the negative
correlation is more clearly seen for cool-core clusters.  This
behavior is in favor of a quasi-hydrostatic cooling picture.

\subsection{Beta-model Fit to Surface Brightness Profiles: Physical
  Basis of Applying Double $\beta$-Model to Cool Core
  Clusters}\label{subset:betafit_sb}

The density profile of the quasi-hydrostatic cool core can be imitated
by the $\beta$-model with a core radius smaller than that of the
initial noncool core, as demonstrated in figure~\ref{fig:density}.
This suggests that either $\beta$-model or double $\beta$-model can be
still fitted to a cool-core cluster, depending also on the available
range of radius.  Observationally, the double $\beta$-model is
frequently used to be fitted to the X-ray data of clusters having
compact cores or cool cores (e.g., \cite{ota04,santos10}).  In this
subsection, we examine whether the conventional single-component or
two-component $\beta$-models can be fitted to the X-ray surface
brightness expected under quasi-hydrostatic cooling.

\begin{figure*}
\begin{center}
\FigureFile(70mm,80mm){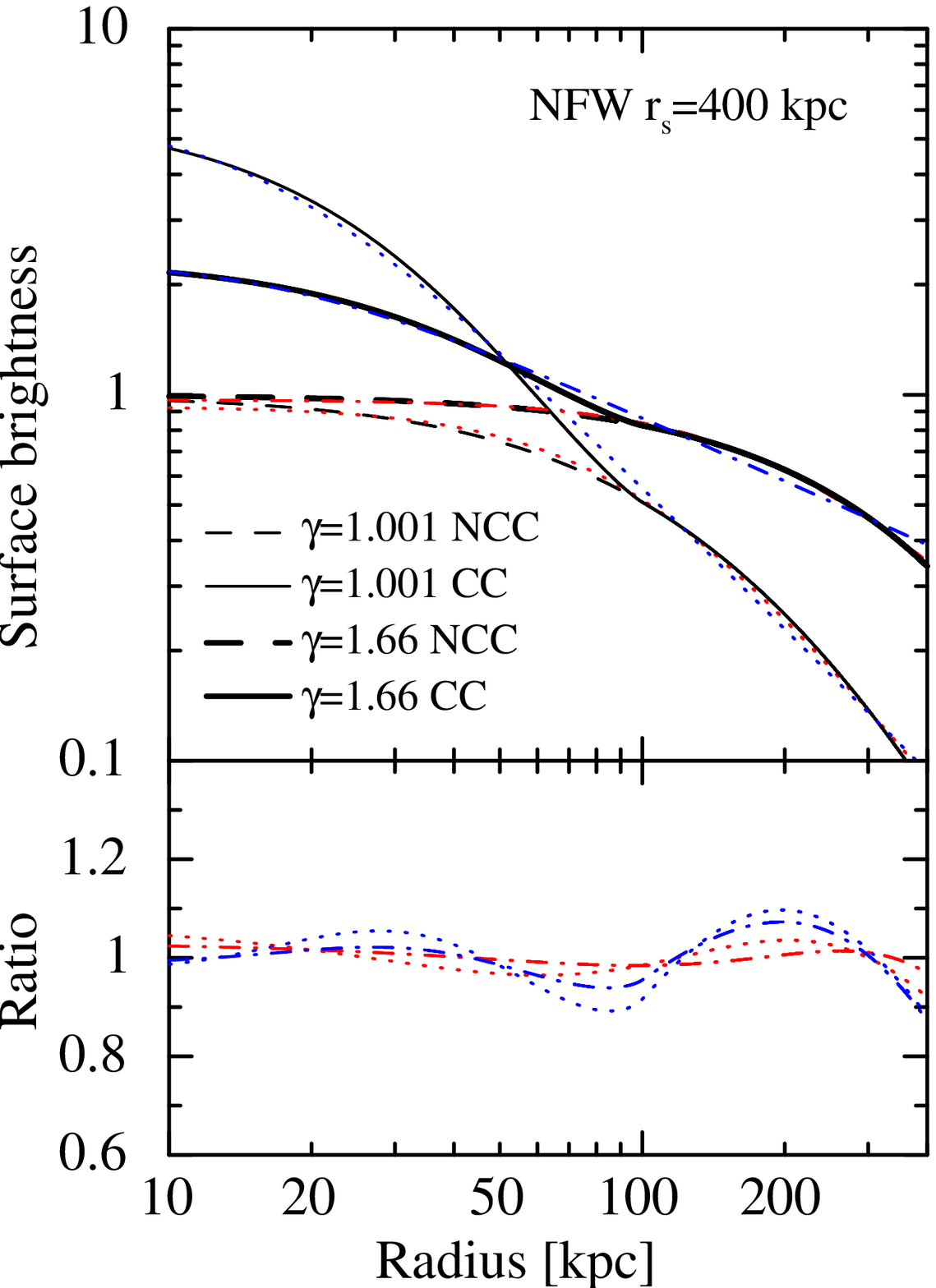}
\FigureFile(70mm,80mm){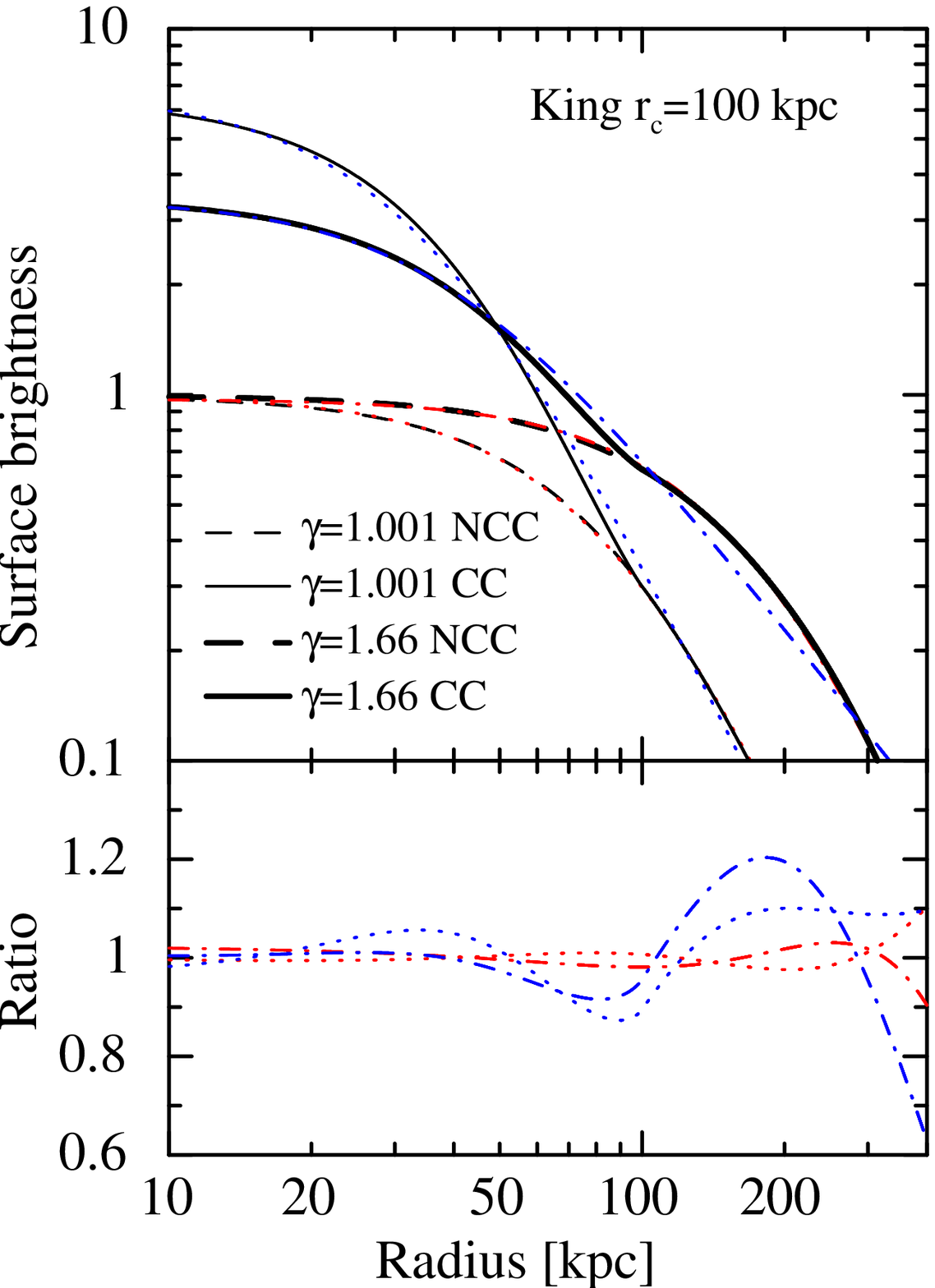}
\end{center}
\caption{Surface brightness of free--free emission calculated for
  quasi-hydrostatic cool cores (solid lines, CC) and the initial
  noncool cores (broken lines, NCC). In the upper panels, the
  best-fit single $\beta$-model are also indicated by the red/blue
  dotted curves for the NCC/CC cases with $\gamma=1.001$ and the
  red/blue dash-dotted curves for the NCC/CC cases with $\gamma=1.66$,
  respectively. In the lower panels, the ratios of the calculated
  surface brightness profiles to the best-fit models are plotted.}
\label{fig:sb}
\end{figure*}

We calculated the surface brightness of free--free emission ($\propto
\rho'^2 T'^{1/2}$ or $\propto \rho^2 T^{1/2}$ of cool or noncool
core).  As can readily be seen in figure~\ref{fig:sb}, the gas core
appears smaller in the quasi-hydrostatic cool core than the initial
noncool one.  We performed the $\beta$-model fitting to the calculated
surface brightness profiles with the form $S(r) = S_0 [ 1+ (r/r_{\rm
  fit})^2]^{-3\beta+1/2}$. A radial range of $5< r/{\rm kpc} < 400$
was used. The best-fit parameters are given in
table~\ref{tab:betafit_sb}.  The surface brightness is enhanced by a
factor of 2.4--6.5, and the core radius becomes smaller by a factor of
$\sim 2.5$--10 for the quasi-hydrostatic cool core in comparison with
the initial noncool core. In the lower panels of figure~\ref{fig:sb},
the ratio of the calculated surface brightness profiles to the
best-fit $\beta$-models are shown. The single-component $\beta$-model
is reasonably well fitted to the NCC case, while it leaves a
systematic deviation from that of the CC case. The chi-squared value,
defined as $\chi^2\equiv\sum [ I(r) - S(r) ]^2 $, is also given in
table~\ref{tab:betafit_sb}. Here, $I(r)$ represents the calculated
surface brightness profile and $\sum$ sums up all data points.

\begin{figure*}
\begin{center}
\FigureFile(70mm,80mm){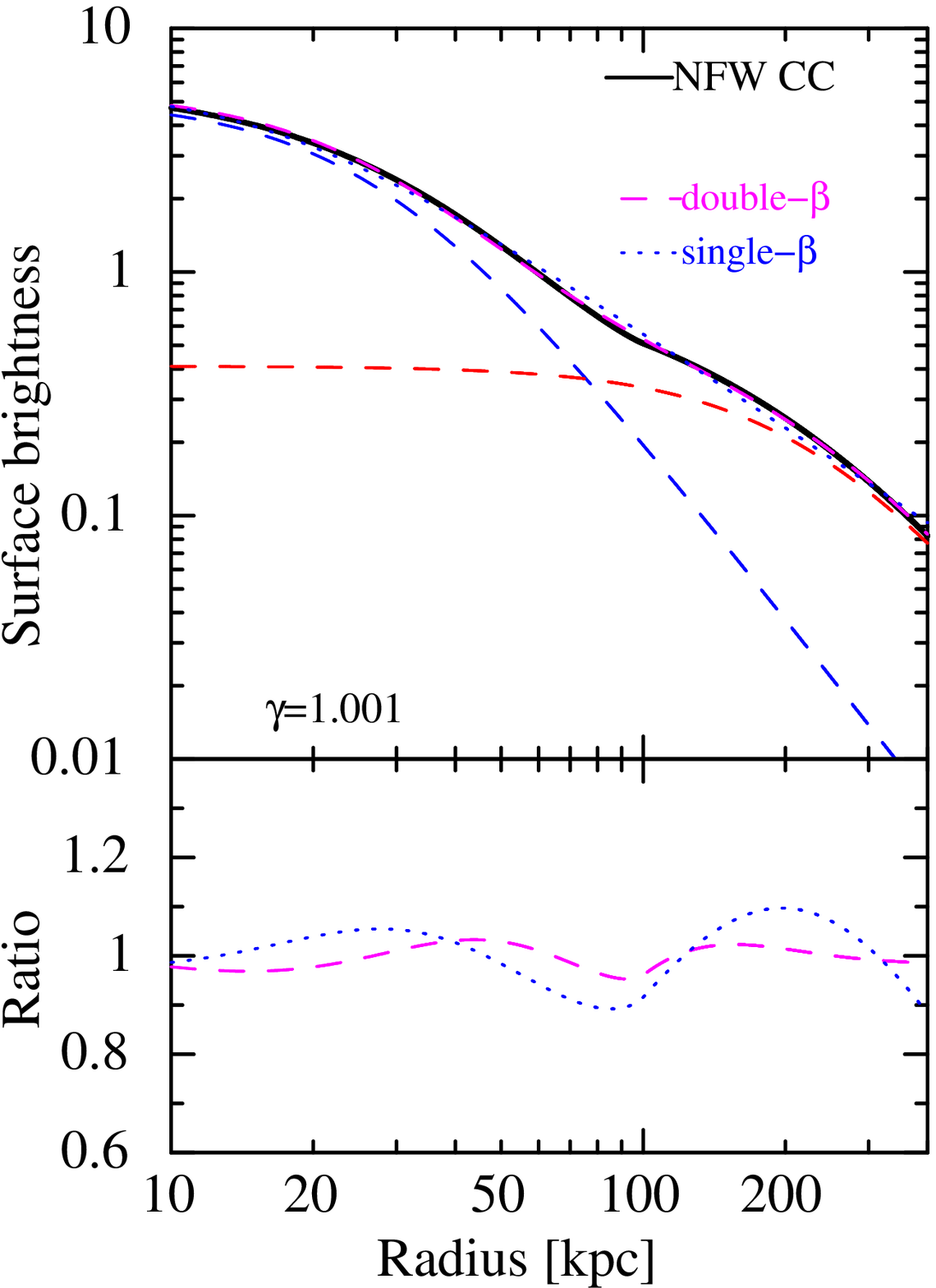}
\FigureFile(70mm,80mm){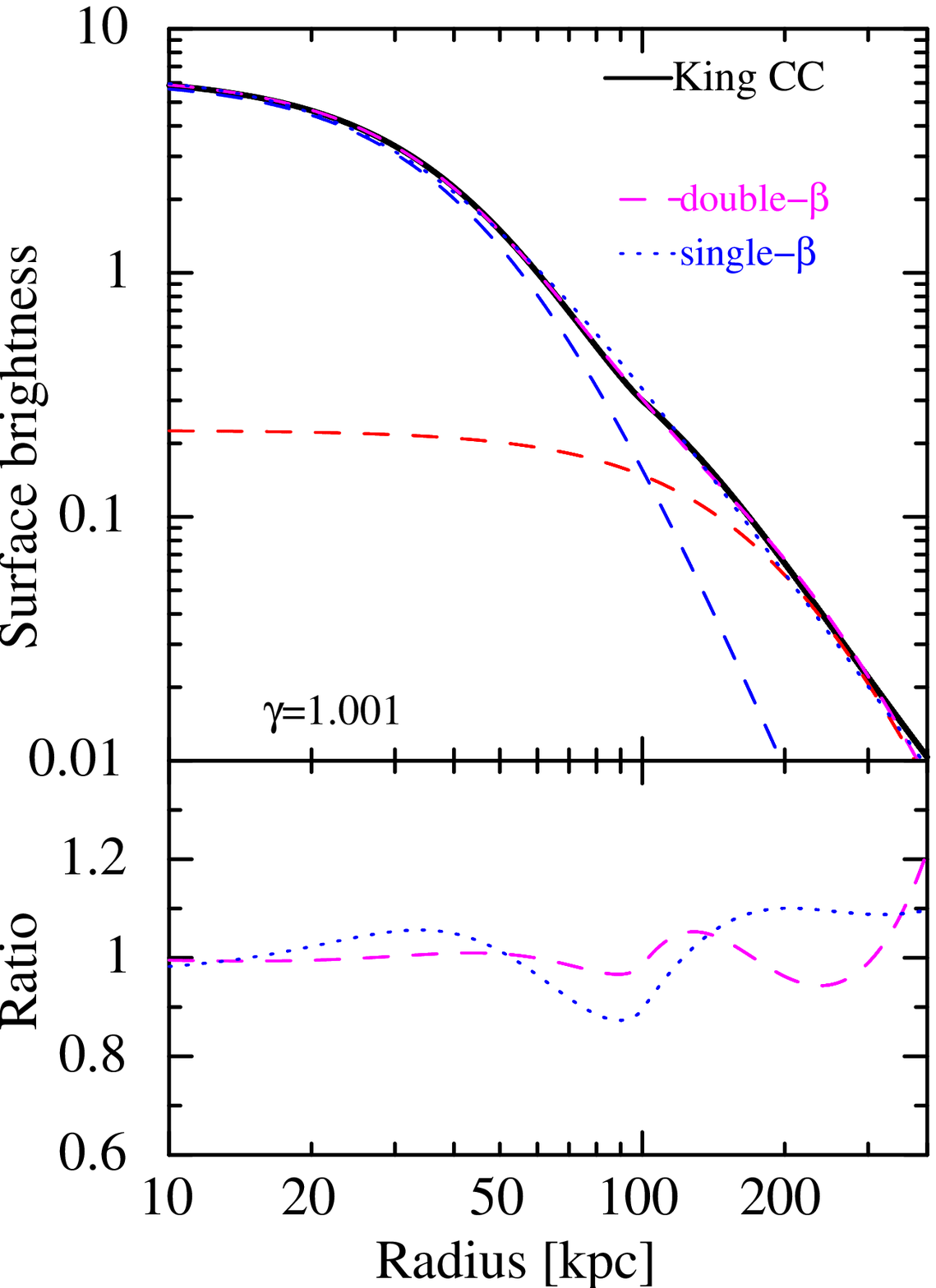}
\FigureFile(70mm,80mm){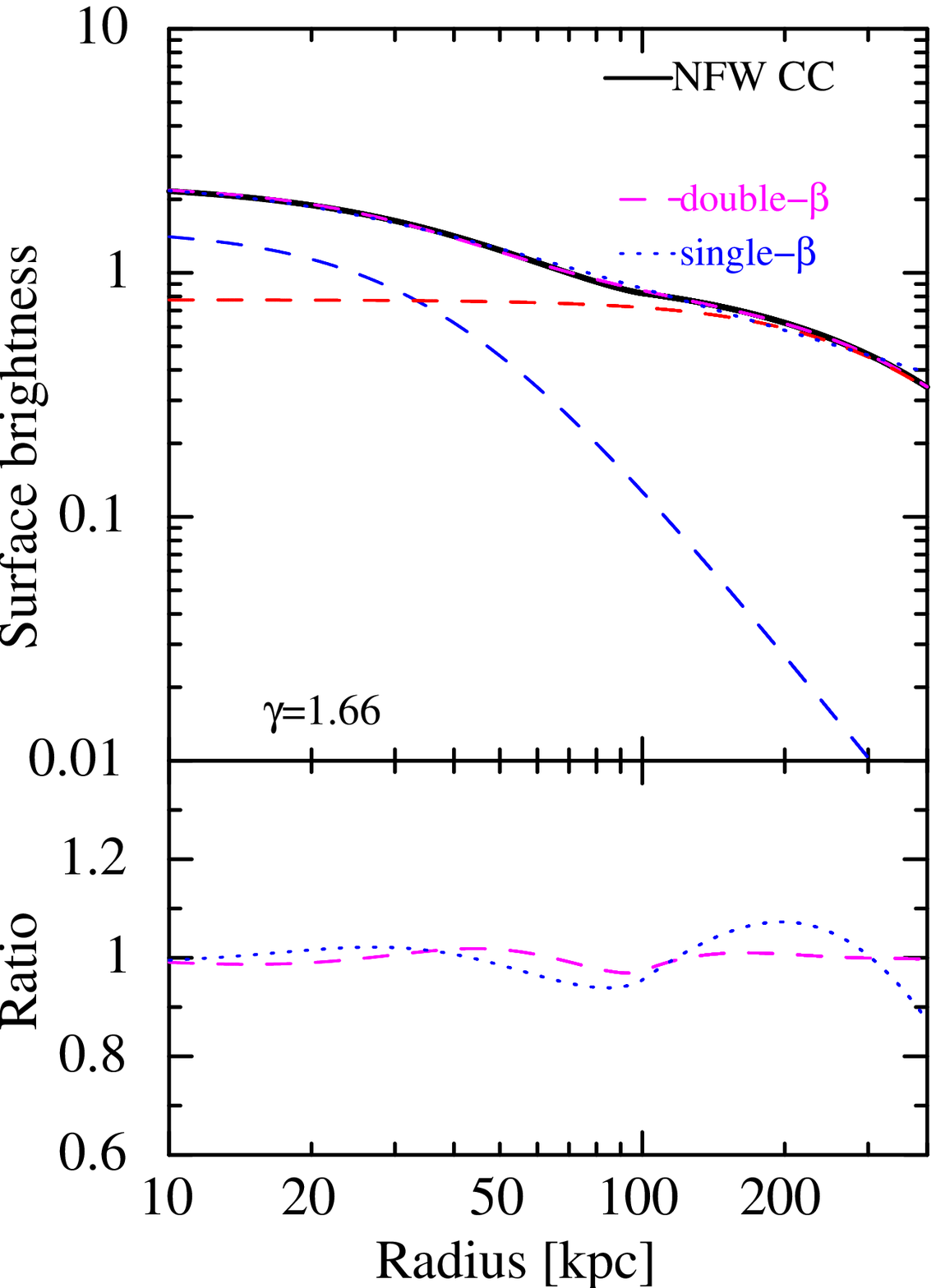}
\FigureFile(70mm,80mm){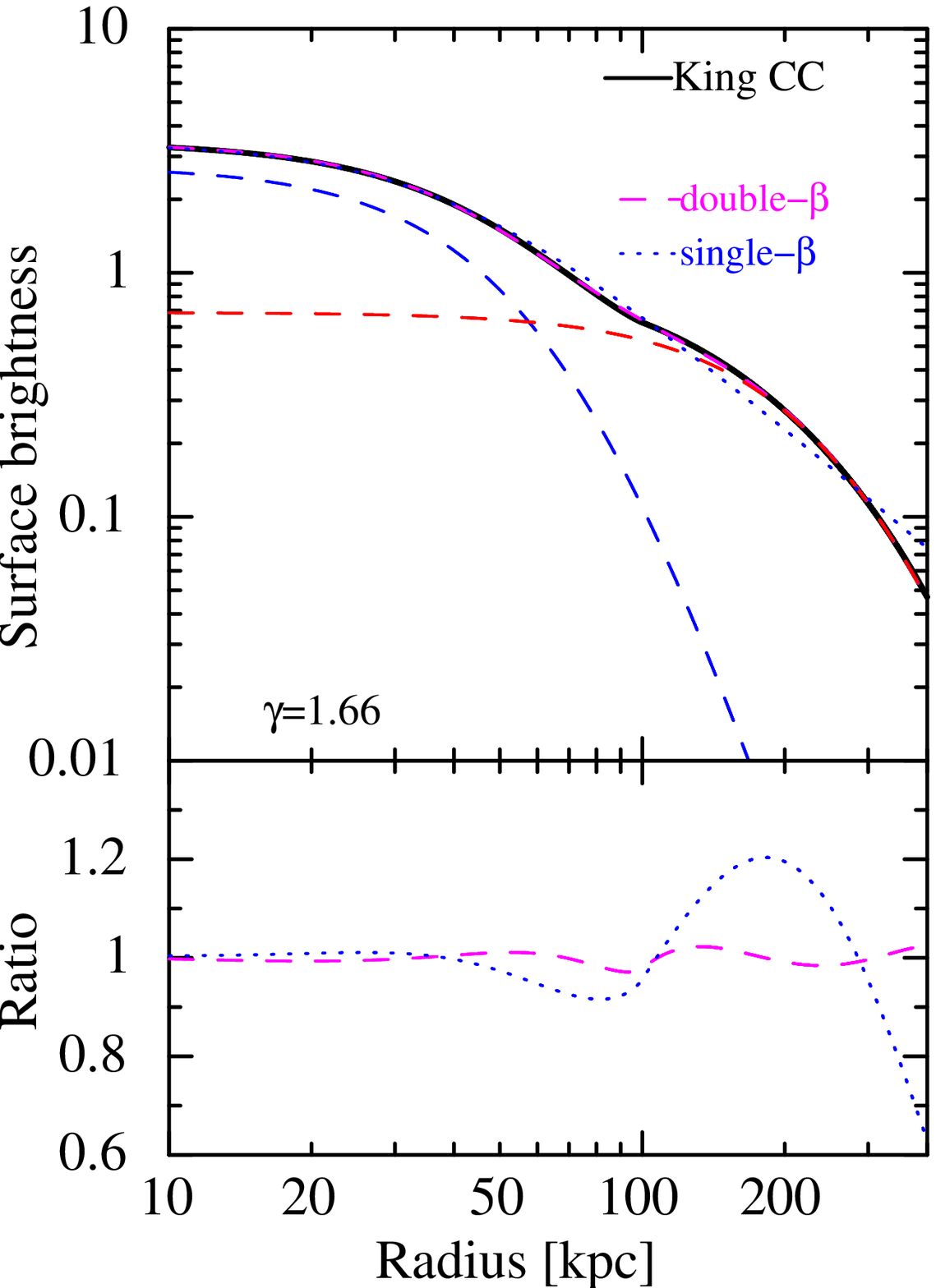}
\end{center}
\caption{Surface brightness of free--free emission calculated for
  quasi-hydrostatic cool cores (solid lines, CC). In the upper
  panels, the best-fit single and double $\beta$-models are
  indicated with the blue dotted and magenta dashed curves.  The
  contributions of two components of the double $\beta$-model are also
  shown with the red and blue curves, respectively.  In the lower
  panels, the ratios of the calculated surface brightness for the
  cool core to the best-fit single and double $\beta$-models 
  are shown with the blue dotted and magenta dashed curves,
  respectively.}
\label{fig:sb_wbeta}
\end{figure*}

Next, the double $\beta$-model, $S(r) = \sum_{i=1}^{2} S_{0,i} [1 +
(r/r_{{\rm fit},i})^2]^{-3\beta_i+1/2}$, is applied to the CC
profiles.  In this fitting, the inner and outer slope parameters are
tied (i.e., $\beta_1 = \beta_2$) because of a large uncertainty of the
inner slope, and capable of varying within 0--2.  The two core radii,
$r_{{\rm fit},1}$ and $r_{{\rm fit},2}$, are restricted to a range of
5--400~kpc.  As can be seen from figure~\ref{fig:sb_wbeta}, the
surface brightness profile of clusters under radiative cooling can be
well approximated by this two-component model.  In fact, the
systematic deviation in the single $\beta$-model fitting is
significantly reduced, and the $\chi^2$ values become smaller in this
case (table~\ref{tab:betafit_sb}).

The above result gives a physical basis for the use of the double
$\beta$-model in analyses of the X-ray surface brightness of cool-core
clusters. We should note that the resulting $r_{{\rm fit},1}$ or
$r_{{\rm fit},2}$ in this analysis is not necessarily equal to the
core radius assumed for the potential distribution, $r_{\rm c}$, or
that expected through the relationship $r_{\rm c} \sim 0.22r_{\rm s}$
\citep{makino98}.

The inner component with a small core size of $r_{{\rm
    fit},1}\sim20-70$~kpc (table~\ref{tab:betafit_sb}) can be ascribed
to radiatively cooling gas at the center.  The present result will
also explain the fiducial emission measure profile obtained by
\citet{arnaud02} as well as the double-peaked core-size distribution
discovered by \citet{ota02,ota04}, whose peak values are $r_{\rm
  c}=50$~kpc and 200~kpc.  \footnote{Later, a similar distribution has
  been shown independently by \citet{hudson10} using the
  high-resolution {\it Chandra} data on a nearby flux-limited sample.
  Notice that the detailed statistics of the cool-core formation seems
  to admit further investigation (e.g., \cite{eckert11} and references
  therein), which is, however, beyond the scope of the present paper.
}

\begin{table*}
\caption{$\beta$-model parameters for surface brightness profiles in the noncool and cool cores.}
\begin{center}
\begin{tabular}{lllllllllll}\hline\hline
Potential & $\gamma$ & & Model & $S_{0,1}$\footnotemark[*]  & $r_{{\rm fit},1}$~[kpc] & $\beta_1$ 
 & $S_{0,2}$\footnotemark[*] & $r_{{\rm fit},2}$~[kpc] & $\beta_2$ & $\chi^2$ \\ \hline
NFW & 1.001 & NCC& single-$\beta$ & 0.93 (1.00) &   96 &    0.44 &    -- &    -- &    --  &    1.29 \\ 
NFW & 1.001 & CC & single-$\beta$ & 5.86  (5.98) &   16 &    0.38 &    -- &    -- &    --  &    7.13 \\ 
NFW & 1.001 & CC & double-$\beta$ & 5.14  &   28 &    0.58 &    0.41 &  237 &  0.58 &    1.61 \\ 
NFW & 1.66   & NCC& single-$\beta$ & 0.97 (1.00) &  191 &    0.37 &    -- &    -- &    --  &    0.41 \\ 
NFW & 1.66   & CC & single-$\beta$ & 2.34 (2.36) &   18 &    0.26 &    -- &    -- &    --  &    6.57 \\ 
NFW & 1.66   & CC & double-$\beta$ & 1.52  &   38 &    0.57 &    0.78 &  400 &   0.57  &    0.40 \\ \hline
King  & 1.001 & NCC& single-$\beta$ & 0.99 (1.00) &   93 &    0.69 &    -- &    -- &    --  &    0.10 \\ 
King  & 1.001 & CC & single-$\beta$ & 6.64 (6.54) &   35 &    0.61 &    -- &    -- &    --  &    5.57 \\ 
King  & 1.001 & CC & double-$\beta$ & 6.19 &   51 &    0.94 &    0.23 &  223 &    0.94 &    0.57 \\ 
King  & 1.66   & NCC& single-$\beta$ & 0.97 (1.00) &  190 &    0.75 &    -- &    -- &    --  &    0.49 \\ 
King  & 1.66   & CC & single-$\beta$ & 3.42 (3.48) &   39 &    0.44 &    -- &    -- &    --  &   17.45 \\ 
King  & 1.66   & CC & double-$\beta$ & 2.74 &   74 &    1.19 &    0.69 &  336 &   1.19 &    0.30 \\ \hline
\end{tabular}
\end{center}
\footnotemark[*] Central surface brightness normalized by that for the noncool core. The value in parenthesis denotes the central surface brightness obtained from the present calculation. 
\label{tab:betafit_sb}
\end{table*}%

\subsection{Implication on the Time Scale of ICM Thermal Evolution}\label{subsec:timescale}

The fraction of cool-core clusters will give another important clue to
understand the formation of cool cores, since it should reflect any
time scale relevant to their evolution.

\citet{akahori06} calculated the probability distribution by
estimating the time during which a cluster would have a certain core
size, to confirm the double-peaked nature of core radius.  They
suggest that the gas rapidly cools after the quasi-hydrostatic
condition breaks.  In other words, relaxed clusters with small-core
radii currently observed are more or less those at the
quasi-hydrostatic cooling stage, which terminates at $\sim \tau_{\rm
  cool}$.  Since $\tau_{\rm cool}$ is estimated to be smaller than the
Hubble time $H_0^{-1} \sim 10$~Gyr for a typical cluster, some heating
is needed to sustain the system; otherwise, it would be unseen in
$\sim$~Gyr after being virialized.  Practically, however, heating due
to mergers is likely invoked in cluster evolution, as also suggested
from a point of view of radio halos (e.g., \cite{rossetti11}).  The
clusters of core radii $> 400$~kpc in the core-size distribution are
possibly attributed to mergers due to their morphology; it should be
noted that quasi-hydrostatic cooling is a picture for a relaxed
gravitational system.  Also other heating/pressure sources, e.g., by
black-hole activity, may work.  In fact, many ``cooling flow''
clusters have central cD galaxies hosting black holes.  Radio
minihalos observed in these clusters may suggest a considerable
contribution of the black holes.

Utilizing the technique of the X-ray fundamental plane, \citet{ota06}
investigated the evolution of X-ray observables for the
distant-cluster sample.  They showed that the principle axis of the
plane is parallel to the $t_{\rm cool}$-axis, suggesting that the
cooling time is a parameter to control the core structure. They also
noted that the rate of radiative-energy loss is likely to be kept
nearly constant for a significant duration of time, even after the
onset of cooling in the relaxed systems.  This indicates that some
steady state is attained for the gas of many small-core clusters.
Those results can be consistently understood within the framework of
the quasi-hydrostatic model.
 
Finally, it should be noted that the present calculation shows the
evolution of the density profile in the cluster core when the initial
NCC profile is given by $\rho(r)/\rho(0)$ in subsection
\ref{subsec:initial}; a different initial condition may result in a
different CC profile.  We ignore the mass of inflow gas through our
calculation. At the very late stage of quasi-hydrostatic cooling,
however, the inflow gas could alter the gravitational potential in the
inner-core region to a steeper profile from the initial equilibrium
one.  This would break the quasi-hydrostatic condition, and eventually
lead to global cooling flow, unless some counter pressure works
against, e.g., by heating the gas.  Further investigation is needed to
obtain a complete view on the thermal evolution of cluster core,
including the possibility of heat input due to mergers, as mentioned
above.

\section{Summary}

\begin{enumerate}
\item We have calculated the density profile of cool core of
  intracluster gas in the quasi-hydrostatic cooling phase
  \citep{masai04}, while assuming that the gas is initially in the
  virial equilibrium state within the dark-matter potential
  represented by the NFW or King distribution, and has a polytropic
  profile with $\gamma=1.001$ or 1.66.
\item In the quasi-hydrostatic cooling phase, the gas density
  increases by a factor of 4--6 at the cluster center, while the
  temperature decreases to about one-third compared with their initial
  polytropic profiles (noncool core, NCC). Accordingly, the core
  radius of the cool core (CC) cluster appears to be more compact than
  the initial polytropic one. This result is consistent with the
  hydrodynamical simulations by \citet{akahori06}.
\item Compared with X-ray observations, it is found that
  the density profile of CC falls between the mid- and high-values of a 
  small-core ($r_{\rm c} < 100$~kpc) when normalized so that the initial-density 
profile has a cooling time equal to 10~Gyr at the cooling radius.
\item The pressure and entropy profiles are derived for the
  quasi-hydrostatic cooling gas, and their CC/NCC profiles are compared
  with the generalized NFW pressure profile for the REXCESS cluster
  sample \citep{arnaud10} and the best-fit entropy profile models to 
  the ACCEPT sample \citep{cavagnolo09}, respectively.
\item The X-ray brightness profiles are calculated for clusters having
  either the NFW or King potential, while assuming free--free
  emission. The resulting CC profiles are found to be well represented
  by the conventional double $\beta$-model including the central
  emission with a compact-core size. This gives a physical basis of
  applying the double $\beta$-model to observed CC clusters. The inner
  component of the double $\beta$-model, which can be ascribed to the
  radiatively cooling gas at the center, will also explain the
  fiducial emission measure profiles \citep{arnaud02} as well as the
  double-peaked core-size distribution \citep{ota02,ota04}.
\item The implication on the time scale of ICM thermal evolution is
  discussed. Since the quasi-hydrostatic stage will terminates at
  $\tau_{\rm cool} \sim {\rm Gyr} < H_0^{-1}$ for a typical cluster,
  some heating is needed to sustain the system. The mass of inflow gas
  is ignored in the present calculation; however, it would alter the
  the potential distribution of the inner core to an even steeper one
  at the late stage of cooling, which would break the
  quasi-hydrostatic balance and lead to global-cooling
  flow. Further investigation is needed to construct a total view concerning 
  the thermal evolution, including the possibility of heat input due
  to mergers.
\end{enumerate}

\bigskip This work was supported in part by a Grant-in-Aid for Young
Scientists by MEXT, No. 22740124 (NO) and by a Grant-in-Aid for
Scientific Research 18540241 from JSPS (KM).

\appendix
\section*{Systematic Error of Central Gas
  Density}\label{appendix:syserr}
\begin{figure}
\begin{center}
\FigureFile(75mm,100mm){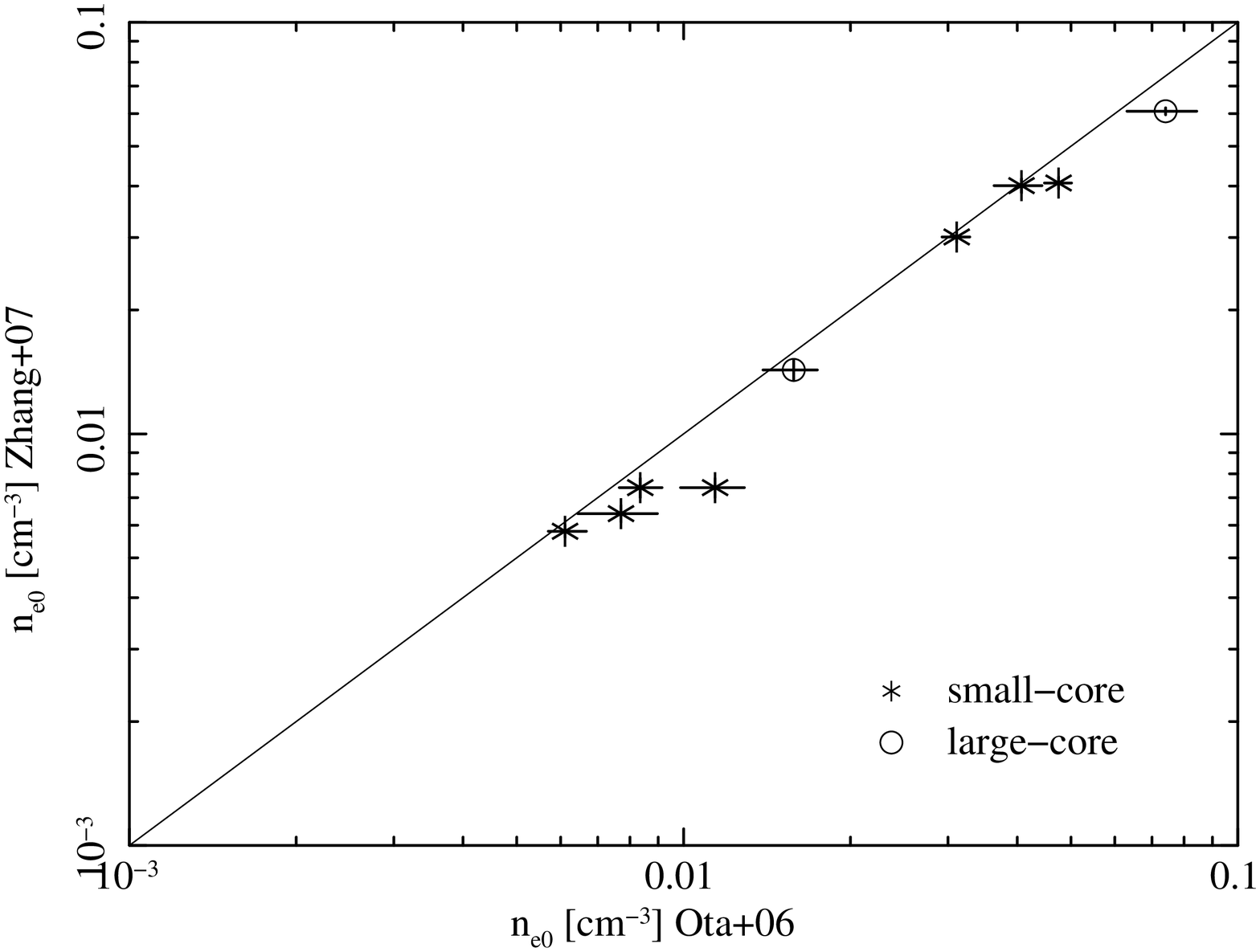}
\end{center}
\caption{Comparison of central gas density between two different
  samples, \citet{ota06} and \citet{zhang07}. The small- and
  large-core clusters are denoted by the asterisks and circles,
  respectively}\label{fig:ne_syserr}
\end{figure}

\citet{ota06} derived the electron density by an isothermal
$\beta$-model fitting to the surface-brightness profiles. Since the
temperature dependence of the X-ray surface brightness is generally
small ($\propto T^{1/2}$), the systematic error of the central gas
density coming from the temperature drop at the cluster center is
expected to be small. Actually, we made a comparison of the central
gas density obtained from the {\it XMM-Newton} observations, having
improved resolutions \citep{zhang07}, and confirmed that the two
samples are in agreement, as shown in figure \ref{fig:ne_syserr}. The
figure includes 7 cool-core clusters as well as 2 noncool-core
clusters. Thus, the evolution of gas-density profile can be further
inferred by comparing the gas densities calculated for the cool and
noncool cores in the present work with the $\beta$-model profiles
presented in \citet{ota06}.

\begin{thebibliography}{}

\bibitem[Akahori \& Masai(2005)]{akahori05} Akahori, T., \& Masai, K.\
  2005, \pasj, 57, 419

\bibitem[Akahori \& Masai(2006)]{akahori06} Akahori, T., \& Masai, K.\
  2006, \pasj, 58, 521

\bibitem[Allen et al.(2001)]{allen01} Allen, S.~W., Schmidt, R.~W., \&
  Fabian, A.~C.\ 2001, \mnras, 328, L37

\bibitem[Arnaud et al.(2002)]{arnaud02} Arnaud, M., Aghanim, N., \&
  Neumann, D.~M.\ 2002, \aap, 389, 1

\bibitem[Arnaud et al.(2010)]{arnaud10} Arnaud, M., Pratt, G.~W.,
  Piffaretti, R., B{\"o}hringer, H., Croston, J.~H., \& Pointecouteau,
  E.\ 2010, \aap, 517, A92

\bibitem[Bulbul et al.(2010)]{bulbul10} Bulbul, G.~E., Hasler, N.,
  Bonamente, M., \& Joy, M.\ 2010, \apj, 720, 1038
  
\bibitem[Bonamente et al.(2006)]{bonamente06} Bonamente, M., Joy,
  M.~K., LaRoque, S.~J., et al.\ 2006, \apj, 647, 25

\bibitem[Cavagnolo et al.(2009)]{cavagnolo09} Cavagnolo, K.~W.,
  Donahue, M., Voit, G.~M., \& Sun, M.\ 2009, \apjs, 182, 12

\bibitem[Chen et al.(2007)]{chen07} Chen, Y., Reiprich, T.~H.,
  B{\"o}hringer, H., Ikebe, Y., \& Zhang, Y.-Y.\ 2007, \aap, 466, 805

\bibitem[Eckert et al.(2011)]{eckert11} Eckert, D., Molendi, S., \&
  Paltani, S.\ 2011, \aap, 526, A79

\bibitem[Fabian(1994)]{fabian94} Fabian, A.~C.\ 1994, \araa, 32, 277 

\bibitem[Hudson et al.(2010)]{hudson10} Hudson, D.~S., Mittal, R.,
  Reiprich, T.~H., Nulsen, P.~E.~J., Andernach, H., \& Sarazin, C.~L.\
  2010, \aap, 513, A37

\bibitem[Jones \& Forman(1984)]{jones84} Jones, C., \& Forman, W.\
  1984, \apj, 276, 38

\bibitem[Makino et al.(1998)]{makino98} Makino, N., Sasaki, S., 
\& Suto, Y.\ 1998, \apj, 497, 555 

\bibitem[Masai \& Kitayama(2004)]{masai04} Masai, K., \& Kitayama, T.\
  2004, \aap, 421, 815

\bibitem[Nagai et al.(2007)]{nagai07} Nagai, D., Kravtsov, A.~V., \&
  Vikhlinin, A.\ 2007, \apj, 668, 1

\bibitem[Navarro et al.(1997)]{navarro97} Navarro, J.~F., Frenk,
  C.~S., \& White, S.~D.~M.\ 1997, \apj, 490, 493

\bibitem[Neumann \& Arnaud(1999)]{neumann99} Neumann, D.~M., \&
  Arnaud, M.\ 1999, \aap, 348, 711

\bibitem[O'Hara et al.(2006)]{ohara06} O'Hara, T.~B., Mohr, J.~J.,
  Bialek, J.~J., \& Evrard, A.~E.\ 2006, \apj, 639, 64

\bibitem[Ota \& Mitsuda(2002)]{ota02} Ota, N., \& Mitsuda, K.\ 2002,
  \apj, 567, L23

\bibitem[Ota \& Mitsuda(2004)]{ota04} Ota, N., \& Mitsuda, K.\ 2004,
  \aap, 428, 757

\bibitem[Ota et al.(2006)]{ota06} Ota, N., Kitayama, T., Masai, K., \&
  Mitsuda, K.\ 2006, \apj, 640, 673

\bibitem[Peterson \& Fabian(2006)]{peterson06} Peterson, J.~R., \&
  Fabian, A.~C.\ 2006, \physrep, 427, 1

\bibitem[Rossetti et al.(2010)]{rossetti11} Rossetti, M., Eckert, D.,
  Cavalleri, B.~M., Molendi, S., Gastaldello, F. \& Ghizzardi, S. \
  2011, \aap, 532, A123

\bibitem[Santos et al.(2008)]{santos08} Santos, J.~S., Rosati, P.,
  Tozzi, P., B{\"o}hringer, H., Ettori, S., \& Bignamini, A.\ 2008,
  \aap, 483, 35

\bibitem[Santos et al.(2010)]{santos10} Santos, J.~S., Tozzi, P.,
  Rosati, P., \& B{\"o}hringer, H.\ 2010, \aap, 521, A64


\bibitem[Snowden et al.(2008)]{snowden08} Snowden, S.~L., Mushotzky,
  R.~F., Kuntz, K.~D., \& Davis, D.~S.\ 2008, \aap, 478, 615

\bibitem[Suto, Sasaki \& Makino (1998)]{suto98} Suto, Y., Sasaki, S., \& Makino,
  N.\ 1998, \apj, 509, 544

\bibitem[Vikhlinin et al.(2006)]{vikhlinin06} Vikhlinin, A., Kravtsov,
  A., Forman, W., Jones, C., Markevitch, M., Murray, S.~S., \& Van
  Speybroeck, L.\ 2006, \apj, 640, 691

\bibitem[Zhang et al.(2007)]{zhang07} Zhang, Y.-Y., Finoguenov, A.,
  B{\"o}hringer, H., et al.\ 2007, \aap, 467, 437

\bibitem[Zhang et al.(2008)]{zhang08} Zhang, Y.-Y., Finoguenov, A.,
  B{\"o}hringer, H., et al.\ 2008, \aap, 482, 451


\end{thebibliography}
\end{document}